\documentclass[twocolumn,showpacs,preprintnumbers,amsmath,amssymb]{revtex4}
\usepackage{graphicx}
\usepackage{dcolumn}
\usepackage{bm}
\usepackage{array}
\usepackage{amssymb}
\usepackage{amssymb,bm,mathrsfs,bbm,amscd}
\usepackage{subfigure}
\usepackage{url}
\usepackage{color}

\begin{document}

\title[]{Dual-induced multifractality in online viewing activity}
\author{Yu-Hao Qin$^{1,2}$},
\author{Zhi-Dan Zhao$^{1,3}$}
\author{Shi-Min Cai$^{1,3,4}$}
\email{shimin.cai81@gmail.com}
\author{Liang Gao$^{2,5}$}
\email{lianggao@bjtu.edu.cn}
\author{H. Eugene Stanley$^{4}$}
\affiliation{$^{1}$Web Sciences Center, School of Computer Science and
  Engineering, University of Electronic Science and Technology of China,
  Chengdu 611731, P. R. China \\
 $^{2}$Institute of Systems Science, School of traffic and
  transportation, Beijing Jiaotong University, Beijing 100044,
  P. R. China \\
 $^{3}$Big Data Research Center, University of Electronic Science and
  Technology of China, Chengdu 611731, P. R. China \\
 $^{4}$Center for Polymer Studies and Department of Physics, Boston
  University, Boston, Massachusetts 02215, USA \\
  $^{5}$Yinchuan Municipal Bureau of Big Data Management and Service, Yinchuan 750011, P. R. China}


\begin{abstract}

\noindent
Although recent studies have found that the long-term correlations relating
to the fat-tailed distribution of inter-event times exist in human activity,
and that these correlations indicate the presence of fractality, the property of
fractality and its origin have not been analyzed. We use both DFA and
MFDFA to analyze the time series in online viewing activity separating from
Movielens and Netflix. We find long-term correlations at both the
individual and communal level, and that the extent of correlation at the
individual level is determined by the activity level. These long-term
correlations also indicate that there is fractality in the pattern of
online viewing. And, we firstly find a multifractality that results from the
combined effect of the fat-tailed distribution of inter-event times
(i.e., the times between successive viewing actions
of individual) and the long-term correlations in online
viewing activity and verify this finding using three
synthesized series. Therefore, it can be concluded that the
multifractality in online viewing activity is caused by both the
fat-tailed distribution of inter-event times and the long-term correlations,
and that this enlarges the generic property of human activity to include not
just physical space, but also cyberspace.

\end{abstract}


\maketitle
\newpage

\section{Beginning}

\noindent
\textbf{To better understand the long-term correlations
and multifractality in human activity, we analyze the time
series in online viewing activity at both the individual and
communal levels via famous DFA and MFDFA methods. We find that
the long-term correlations at both the individual and communal
level are generic to human activity, and at
the individual level the extent of correlation is determined by the
activity level. These long-term correlations suggest the fractal
pattern in online viewing activity. We further find a multifractality
that results from the combined effect of the fat-tailed distribution of
inter-event times (i.e., the times between successive viewing actions
of individual) and the long-term correlations, which is verified by
using synthesized series and surrogate methods. These empirical
results enlarge this generic property of human activity to include not
just physical space, but also cyberspace.}

\section{Introduction}

\noindent
It is difficult to characterize and understand complex systems because
splitting a complex system into simpler subsystems changes its dynamical
properties \cite{Kantelhardt2009}. Thus researchers focus on macroscopic
properties, e.g., analyzing a time series in which the behavioral
evolution of a complex system is characterized by output records
restricted by time scale. Output records from real-world complex
systems, e.g., stock price fluctuations \cite{Stanley1995}, heart rate
variations \cite{Peng1993,Liebovitch1999a,Ivanov1999,Ivanov2001,Ivanov2009}, and inter-spike intervals
\cite{Liebovitch1999b,Liebovitch2001,Bedard2006}, usually follow a
non-Gaussian probability density function (PDF) and involve fractal
dynamics.

Human activity is itself a complex system. Time series analysis, e.g.,
detrended fluctuation analysis (DFA) \cite{Peng1994,Bunde2000,Kantelhardt2001},
has recently discovered several macroscopic properties in human activity.
For example, a periodic pattern has been found in such human activities as Internet surfing
\cite{Gonccalves2008}, online game logins \cite{Jiang2010}, task
submissions to a Linux server \cite{Baek2008}, and e-commerce purchases
\cite{Dong2013}. Long-term correlations found in many physical,
biological, economic, and ecological systems
\cite{Peng1993,Peng1992,Koscielny-Bunde1998,Makse1995,Makse1998,Liu1999,Cai2006,Cai2009,Rozenfeld2008,Rybski2010EPL}
have also been found in human interactive patterns, and these long-term
correlations become strong as the human activity level increases
\cite{Rybski2009}.

Rybski et al. \cite{Rybski2011,Rybski2012} investigate human
communication in a social network, find a relation between long-term
correlations and inter-event clustering, and provide a model to explain
these correlations. They show that at the individual level the long-term
correlations in a time series of events is determined by the power-law
distribution of the inter-event times (the ``Levy correlations'' in
Ref.~\cite{Rybski2012}), but that at the communal level they are a
generic property of the system caused by interdependencies among
community activities. Zhao et al. \cite{Zhao2012} analyze the time
series of inter-event times in online viewing activity, find that the long-term
correlations (i.e., memory) are restricted by the activity level, and
show an abnormal scaling behavior associated with long-term
anticorrelations caused by the bimodal distribution of inter-event
times. Kivel\"{a} and Porter \cite{Kivela2014} systematically estimate
the inter-event time distribution from finite observation periods in
communication networks.

Although these long-term correlations imply the existence of fractality
in these time series of human activity, two unanswered questions remain:
(i) what category of fractality applies in a time series of human
activity and (ii) what is the origin of the fractality? Using the
Internet technology, we examine the time series in the human activity of two movie
viewing websites, Movielens and Netflix, analyze the long-term
correlations, and find fractality. At the individual level, we apply
DFA to time series of records composed by users with the same activity level
and to corresponding shuffled time series in which each user preserves the inter-event
times. Long-term correlations become strong as activity level
increases. Because the distributions of inter-event times at different
activity levels do not follow a power law, there is a trivial difference
between the Hurst exponents \cite{Hurst1951,Hurst1956} of the original
and the shuffled time series. The empirical result differs somewhat from
that found in human communication activity \cite{Rybski2012}. At the
communal level we similarly analyze the time series of records
aggregated from all user activities and find a stronger long-term
correlations with Hurst exponents, approximatly 0.9 and 1.0 for Movielens
and Netflix, respectively.

To more accurately categorize the fractality and understand its origin,
we use multifractal detrended fluctuation analysis (MFDFA) and probe the
singularity spectrum. We find a dependence between the generalized Hurst
exponent and $q$-order statistical moments that indicates
multifractality in the time series of records at the communal
level. Although multifractality remains after the time series are
shuffled, changes occur in the value of the generalized Hurst
exponent. One result is obviously suggested by the singularity
spectrum. Reference~\cite{Rybski2012} hypothesizes that multifractality
relates with the broad PDF of inter-event times and the long-term correlations
\cite{Kantelhardt2002}. Because this hypothesis is verified by our empirical
results and our synthesized series, we conclude that multifractality exists in human
viewing activity and that the combined effect of the fat-tailed distribution of inter-event times
and long-term correlations causes such multifractality.

\section{Data}

\begin{figure}[!t]
\centering
\includegraphics[width=0.45\textwidth]{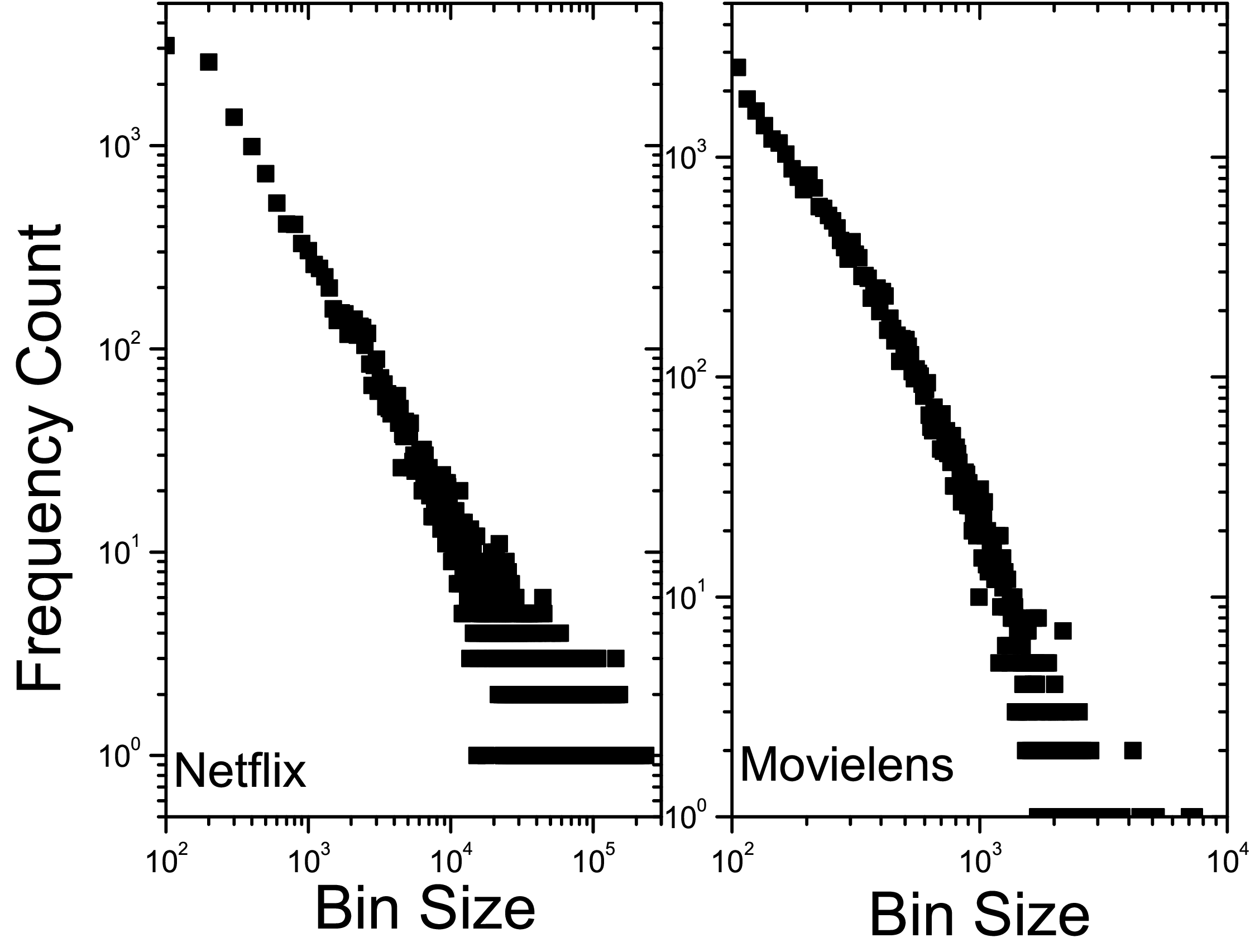}
\caption{\label{fig:timeSeries} The PDFs of user activity levels for
Netflix and Movielens at a log-log scale. They both show a fat-tailed distribution, suggesting
the hierarchical user activities. \label{activity}}
\end{figure}

\noindent
The experimental datasets, released by Movielens and Netflix,
record views and ratings to movies at a given time. Each user's account
is anonymous with a hash tag done by Movielens and Netflix. The total
users are approximate 7,000 for Movielens and 17,774 for Netflix, respectively.
The users' activity levels are hierarchically distributed both in Movielens and Netflix, as shown
in Fig.\ref{activity}. We filter datasets according to user's activity level, $M \geq
55$ (see definition in Sec. 4). Herein, we set $M \geq 55$ to guarantee
that there are abundant records for converting into time series in a long
duration. We finally obtain 26,884 users (38.4$\%$ of
the total users) and 10,000,054 records for a duration of 4,703 days
(nearly 13 years) for Movielens, and 17,703 users (99.6$\%$ of the total
users) and 100,477,917 records for a duration of 2,243 days (nearly 6
years) for Netflix. Although the Movielens records begin at its creation
date, and are thus noisy, the size of both filtered user datasets
exceeds $10^5$.

To convert these records into time series, we introduce two
variables: the records per day of a single user $x(t)$ and the records per
day of the entire community $x_{\rm tot}(t)$. Note that the number of
records quantifies the number of movies viewed by a single user $x(t)$,
which is constrained by the number of hours in a day. We use these time
series in our subsequent analysis. Figures~\ref{fig:timeSeries}(a) and (b)
show the viewing actions of two typical users when they comment movies in Movielens and
Netflix, respectively. Figures.~\ref{fig:timeSeries}(c) and (d) show the corresponding time series at
the individual level. Figures.~\ref{fig:timeSeries}(e) and (f) show the time series at the communal
level. In Fig.~\ref{fig:timeSeries} we can also observe the clusters of activity records
that suggest the burstiness occurring in these online viewing activities.

\begin{figure}[!t]
\centering
\includegraphics[width=0.45\textwidth]{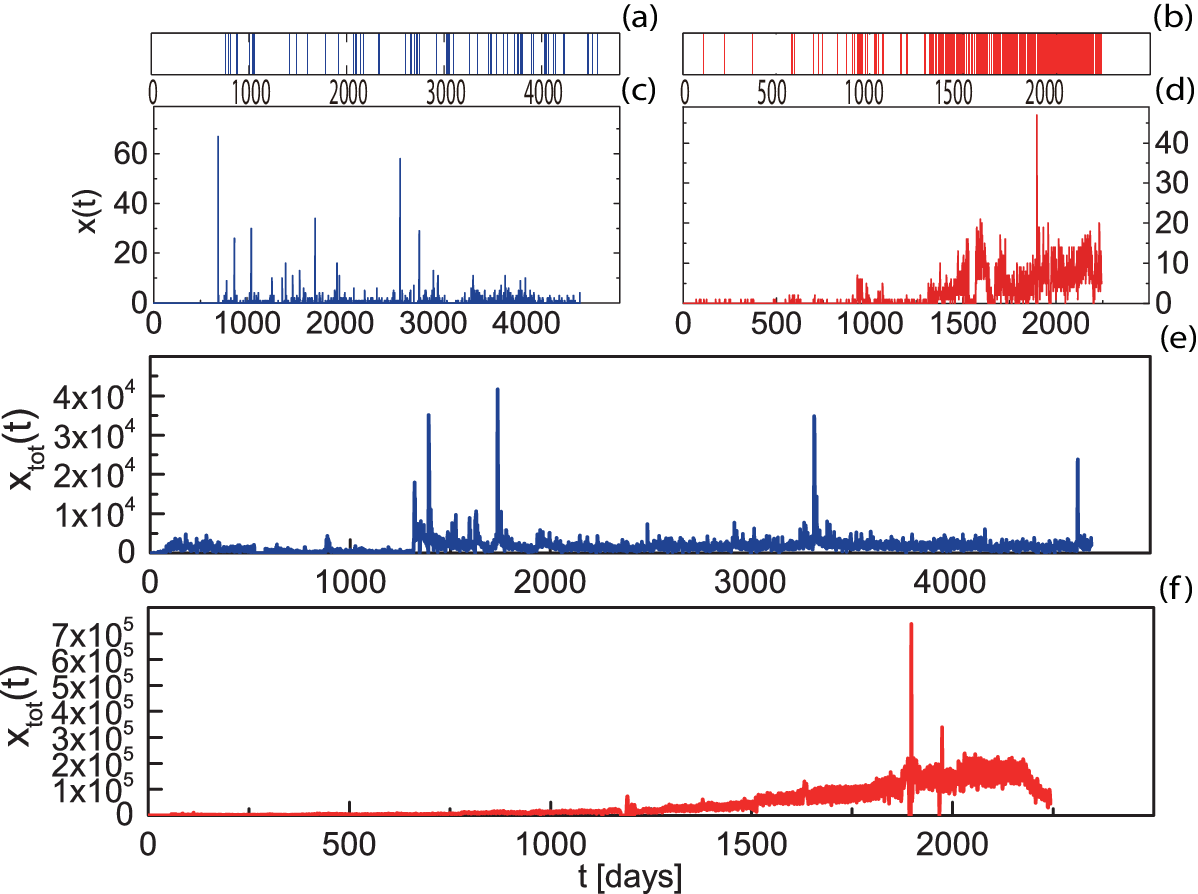}
\caption{\label{fig:timeSeries} (Color Online) A visual illustration of
  activity records of typical users and corresponding time series at both
  individual and communal levels. (a) and (b) indicate the viewing actions of
  two typical users, user23172 ($M=1308$) and user12228 ($M=5606$) when they
  comment movies in the websites. (c) and (d) show corresponding time series
  for these two typical users. (e) and (f) represent time series
  for whole community. The dark blue and red lines denote
  Movielens and Netflix, respectively.}
\end{figure}

\section{Method}

\subsection{Detrended fluctuation analysis}

\noindent
The DFA is a proven method for measuring long-term correlations of time series
\cite{Peng1994,Bunde2000,Kantelhardt2001}, and is less sensitive to
the additional detrending process \cite{Xu2005,Shao2012}, nonstationarities \cite{Chen2002},
noise \cite{Chen2005}, missing data \cite{Ma2010,Xu2011}, etc.
To keep our description self-contained, we briefly introduce the steps of this method as follows:

\begin{itemize}

\item[{(i)}] Calculate the profile $Y(t')$ of time series $x(t)$,
\begin{equation}\label{DFA_profile}
Y(t') = \sum_{t=1}^{t'}x(t)-\langle x(t)\rangle, t' = 1, ..., N.
\end{equation}

  Divide $Y(t')$ into $N_s$ non-overlapping segments of length
  $s$ in increasing order. Because $N$ is often not equal to the
  product of $s$ and $N_s$ [i.e., ($N_s=\lfloor \frac{N}{s} \rfloor$)],
  suggesting that the last segment of $Y(t')$ is overlooked, we divide
  $Y(t')$ from the opposite direction in order to incorporate the entire
  $Y(t')$. Thus there are $2N_s$ different segments. We sample the value
  of $s$ from the logarithmic space, $s = \frac{N}{2^{int(\log_2^N-2)}},
  ..., \frac {N}{2^3}, \frac {N}{2^2}$, which maintains the smoothness
  of the curve between $F(s)$ and $s$.

\item[{(iii)}] Given $s$, the profile $Y(t')$ in each segment is
  detrended separately. The least-square fit is used to determine the
  $\chi^2$-functions for each segment, such as for $v = 1, 2, ..., N_s$,
\begin{equation}\label{DFA_flucDivi1}
F^2(v,s) = \frac{1}{s}\sum_{j=1}^s[Y((v-1)s+j)-\omega_v^n(j)]^2,
\end{equation}
and for $v = N_s+1, ..., 2N_s$,
\begin{equation}\label{DFA_flucDivi2}
F^2(v,s) = \frac{1}{s}\sum_{j=1}^s[Y((N-v-N_s)s+j)-\omega_v^n(j)]^2,
\end{equation}
where $w_v^n$ is the $n$-order polynomial fitting of segment $v$.

\item[{(iv)}] Calculate the fluctuation function,
\begin{equation}\label{DFA_sum}
F(s) = \left[\frac{1}{2N_s}\sum_{v=1}^{2N_s}F^2(v,s)\right]^{\frac{1}{2}} \sim s^H,
\end{equation}
where $H$ is the Hurst exponent. The value of $H$ measures the long-term
correlation of time series. It indicates a long-term anticorrelation for
$0< H< 0.5$, no correlation for $H=0.5$, and a long-term correlation for
$H>0.5$.

\end{itemize}

\subsection{Multifractal detrended fluctuation analysis}

\noindent
DFA gives us the long-term correlations of time series, which indicates
its fractality. To further analyze this fractality and its origin, we
modify DFA and introduce MFDFA \cite{Ivanov1999,Ivanov2001,Kantelhardt2002,Movahed2006,Lim2007,Ivanov2009,Ludescher2011},
where equation (\ref{DFA_sum}) is modified to become
\begin{equation}\label{MFDFA_sum}
F(s) = [\frac{1}{2N_s}\sum_{v=1}^{2N_s}[F^2(v,s)]^{q/2}]^{\frac{1}{q}}
\sim s^{H(q)}.
\end{equation}
Here $H(q)$ is the generalized Hurst exponent. When a time series is
monofractal, $H(q)$ is independent of $q$, when it is multifractal,
$H(q)$ is dependent on $q$. This multifractality is caused by such key
factors as long-term correlations and PDF of inter-event times. To determine the origin of
the multifractality, we randomly shuffle the time series to reduce
long-term correlations but preserve the PDF, and once again apply
MFDFA. If the PDF is the only source of the multifractality, it will be
reserved in the shuffled time series. If the long-term correlations are
the only source, it will disappear. If the long-term correlations and PDF
both affect the time series, the multifractality will remain but the
value of $H(q)$ will change.

The singularity spectrum $f(\alpha)$ provides a clearer way of
characterizing a multifractal time series. The horizonal span of
$f(\alpha)$ quantifies multifractality. A narrow $f(\alpha)$ indicates a
monofractal time series, and a wide $f(\alpha)$ indicates a multifractal
time series. To determine its analytical relationship to $\alpha$, we
introduce Renyi exponent $\tau(q)$ \cite{Meakin1987,Peitgen2004} using
the equation
\begin{equation}\label{tau_q}
\tau(q) = qH(q)+1.
\end{equation}

Applying the Legendre transformation \cite{Tulczyjew1977} gives us the
relation between $f(\alpha)$ and $\alpha$
\begin{equation}\label{alpha_tau}
\alpha = \tau'(q) \ \ and \ \ f(\alpha) = q\alpha-\tau(q),
\end{equation}
or equivalently [using Eq.~(\ref{tau_q})],
\begin{equation}\label{alpha_h}
\alpha = H(q)+qH'(q) \ \ and \ \ f(\alpha) = q[\alpha-H(q)]+1.
\end{equation}


\section{RESULTS}

\subsection{Long-term correlation in individual activity}

\noindent
To sort users by activity level, we define $M_i$ to be the total records
of a single user $i$ ($M_i=\sum_{i=1}^{N} {x_i(t)}$, where $N$ is the
length of the series) and convert $M_i$ into a logarithmic scale,
$L_i=\lfloor (\ln M_i) \rfloor$. Here the range of $L$ is from $4$ to
$8$ in Movielens and from $4$ to $12$ in Netflix.

\begin{figure}[!t]
\centering
\includegraphics[width=0.45\textwidth]{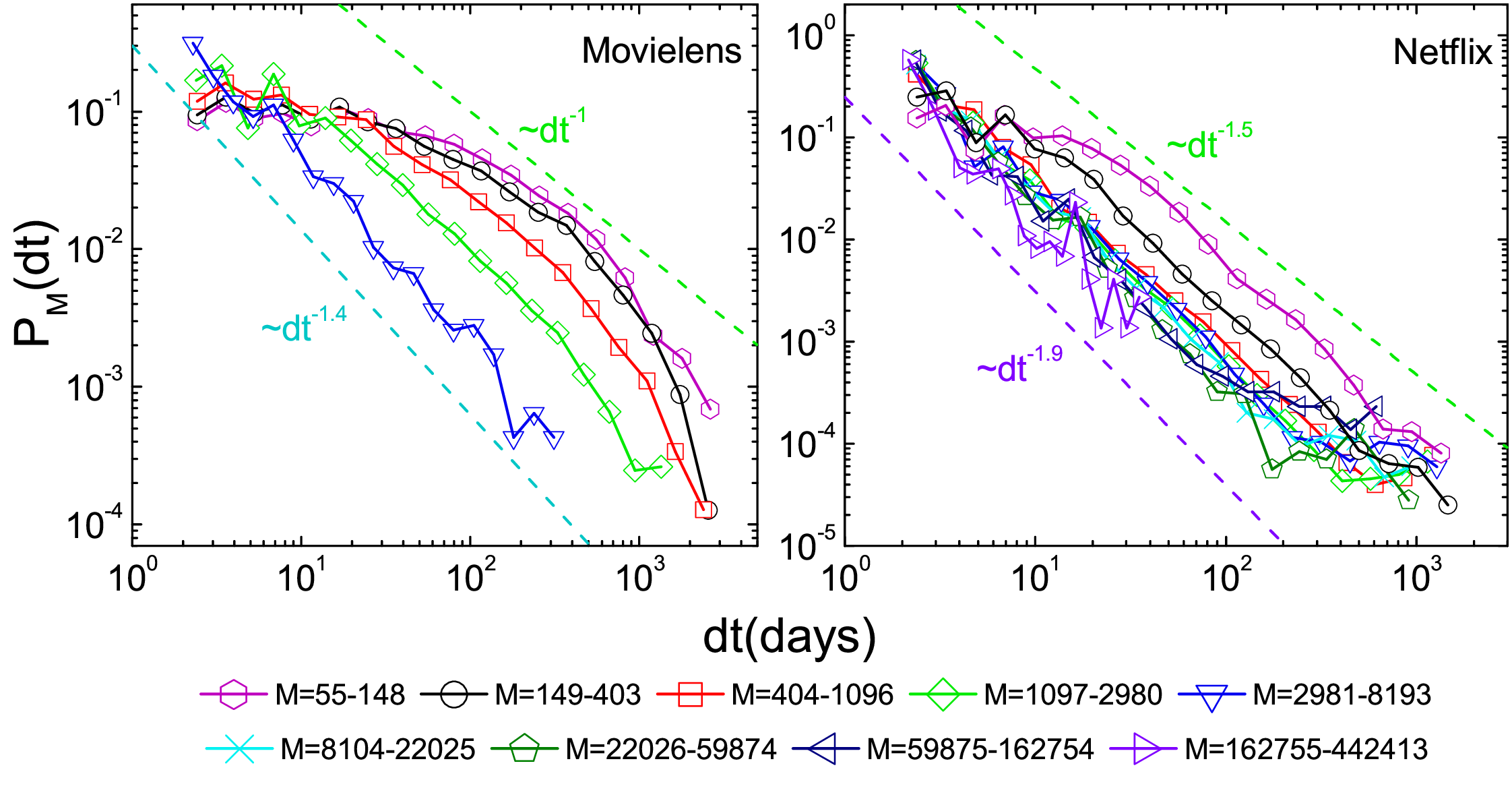}
\caption{\label{fig:timeInterval_level}(Color online) Inter-event time
  distribution at different activity levels.  The left and right panels
  respectively indicate Movielens and Netflix. The dash lines are guide
  for power-law distributions. The fat tails are both found in
  inter-event time distributions for Movielens and Netflix, which
  suggests the burstiness of online viewing activity.}
\end{figure}

\begin{figure}[!t]
\centering
\includegraphics[width=0.45\textwidth]{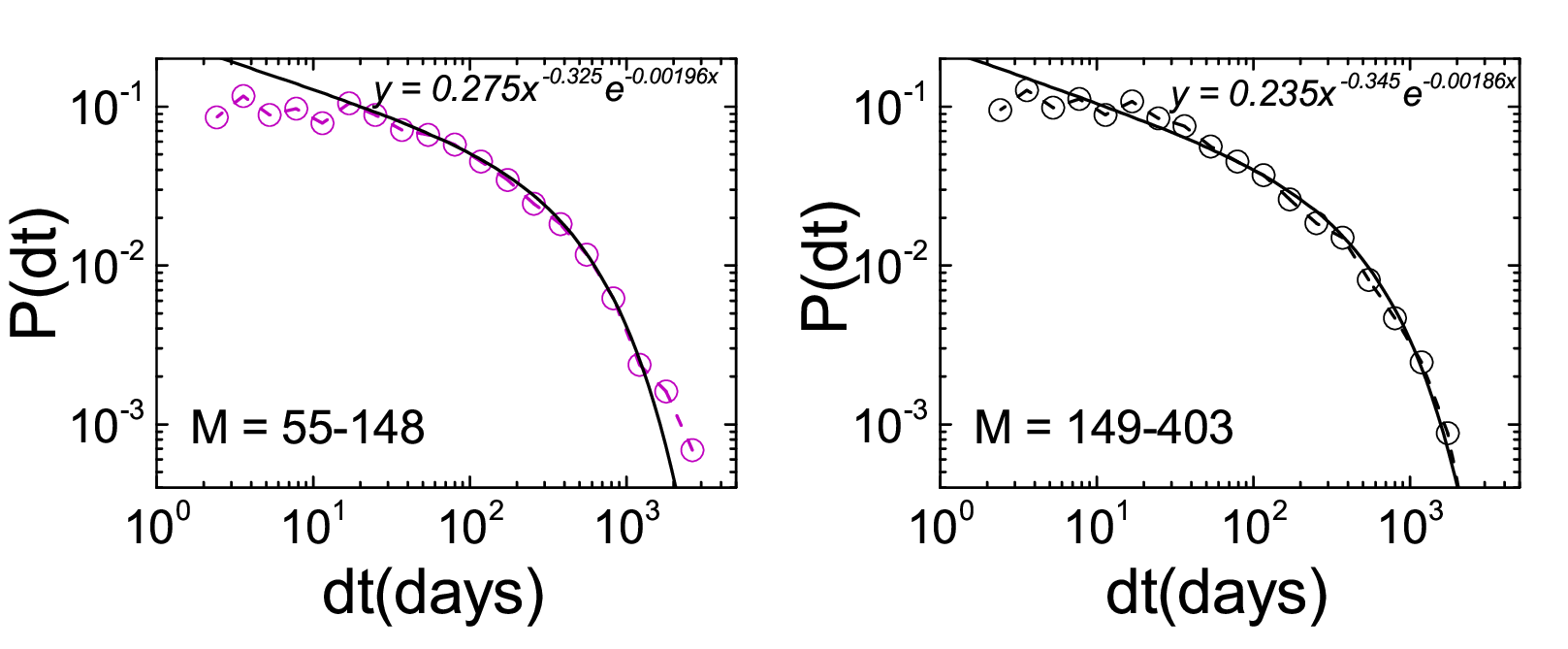}
\caption{\label{fig:small_level} (Color online) Inter-event time
  distributions of users in Movielens whose activity levels are
  respectively $L=4$ and $L=5$. Compared with exact power-law
  distributions, they can be fitted by exponential cut-off power-law
  distributions via least squared estimating method.}
\end{figure}

According to logarithmic activity levels,
we firstly present the distributions of inter-event times in Fig.~\ref{fig:timeInterval_level},
where the left and right panels indicate Movielens and Netflix, respectively.
As shown in Fig.~\ref{fig:timeInterval_level}, both of them show fat tails, which
suggests the burstiness occurring in online viewing activity~\cite{Barabasi2005}. More concretely,
for these users with lower activity levels (e.g. $L<6$), their inter-event times are not
exactly power law distributed. For example, in Fig.~\ref{fig:small_level},
the inter-event times of users with activity levels $L=4$ and $L=5$ in Movielens apparently
follow exponential cut-off power-law distributions via least squared estimating method.
While for these users with larger activity levels (e.g. $L>7$), their distributions of inter-event times
are approximately power law. Thus, the power-law distribution is not the
only type to characterize the fat tail of inter-event clustering (i.e., burstiness),
and this differs somewhat from the empirical data in human communication
\cite{Rybski2012} and stock trading \cite{Ivanov2004,Ivanov2014}.



\begin{figure}[!t]
\centering
\includegraphics[width=0.45\textwidth]{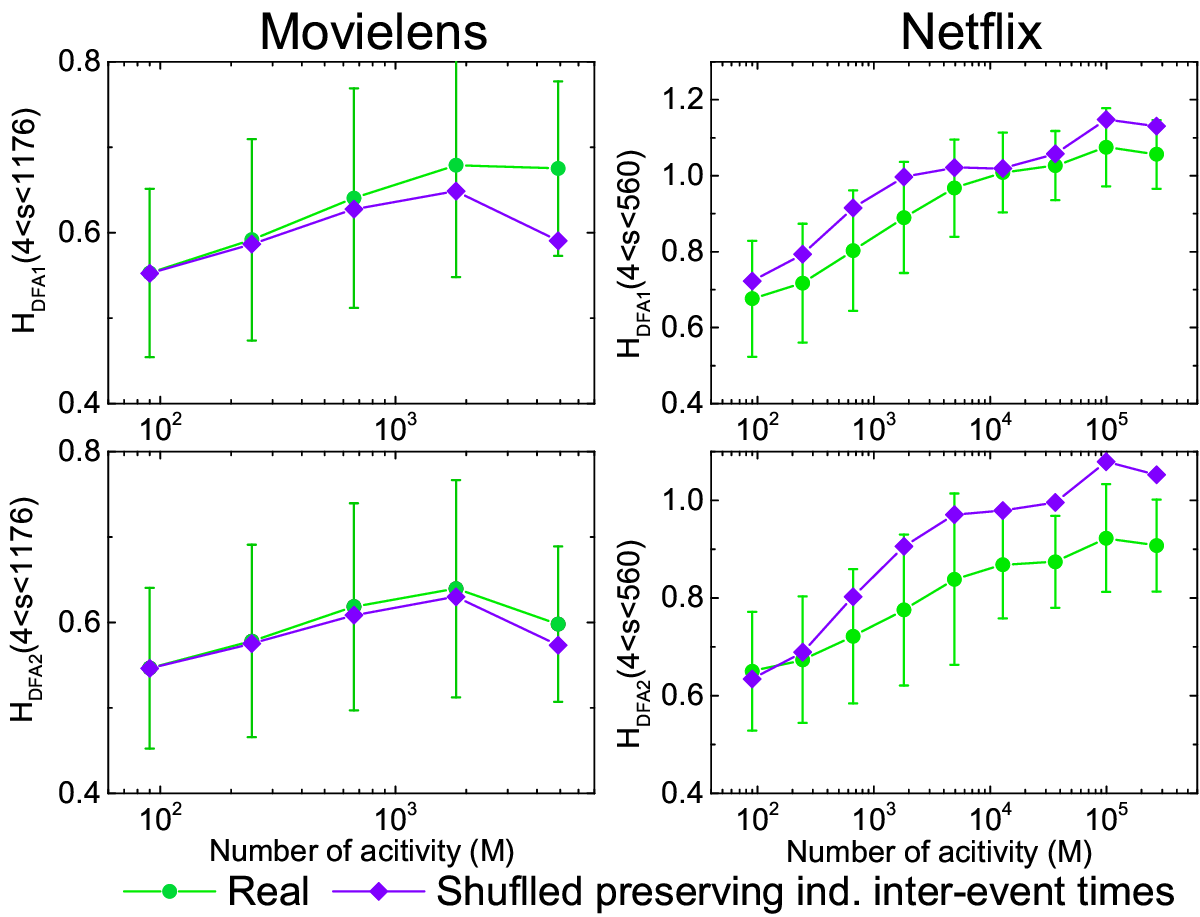}
\caption{\label{fig:DFA_level}(Color online) Average Hurst exponents as a
  function of activity level for Movielens and Netflix. The results
  obtained from original time series of records and shuffled ones are
  respectively plotted with green circles and blue square. With the
  increase of activity levels, the long-term correlations become
  stronger. Moreover, the trivial difference between them reveals the
  long-term correlations having a potential relation with fat-tailed
  inter-event time distribution. The error bar is the deviation of Hurst
  exponents from users at a same activity level.
   }
\end{figure}

Because the burstiness of online reviewing activity (or fat tailed inter-event time distribution) potentially
suggests that the time series of records have long-term correlations \cite{Rybski2012},
we use DFA to calculate the Hurst exponent in each time series of single user.
Note that the least square estimating method is applied for fitting trend,
and the F-statistic test confirms the significant of fitting results (see more in Appendix B)
These Hurst exponents are then scaled according to the user activity level and averaged.
Figure~\ref{fig:DFA_level} shows that the average Hurst exponents as a function of activity levels
are greater than 0.5, and they aren't strictly restricted by order of DFA.
Thus, it can be claimed that the long-term correlations exist in these time series
of records at the individual level. It also worthy to be noted that
there is an approximately positive relation between Hurst exponent
and activity level both for Movielens and Netflix, similar to that in
the traded stock market and communication activity \cite{Eisler2006,Eisler2008,Rybski2012}.
In additional, there is also a trivially
different extent of long-term correlation between Movielens and Netflix.
We assume that this difference is to some extent caused by diverse individual
activity pattern in Movielens and Netflix. The commercial website, Netflix, more easily urges users
to form the cluster of consecutive occurred viewing actions and enhance long-term correlations.



We have found that the long-term correlations and fat-tailed
inter-event time distribution both exist in online viewing activity
from Netfilx and Movielens. To further analyze the relation
between the long-term correlation and inter-event time
distribution, we shuffle the time series of records but preserve
the distribution of inter-event times for each user.
The procedure is shown as follows: (i) extract inter-event times of each user;
(ii) shuffle the extracted data; and (iii) keep the first time stamp constant and rebuild the time
series of records using the shuffled data.

We reuse the DFA to obtain the Hurst exponents of the new time series.
Figure~\ref{fig:DFA_level} shows that they differ only trivially
from those of original data, indicating that at the individual level the
long-term correlations of time series of records is associated with the
fat-tailed distribution of the inter-event times. Because the
inter-event times are not strictly power-law distributed, we cannot
infer the long-term correlations from a Levy correlation \cite{Rybski2012}.


\subsection{Long-term correlation in community activity}

\begin{figure}[!t]
\centering
\includegraphics[width=0.45\textwidth]{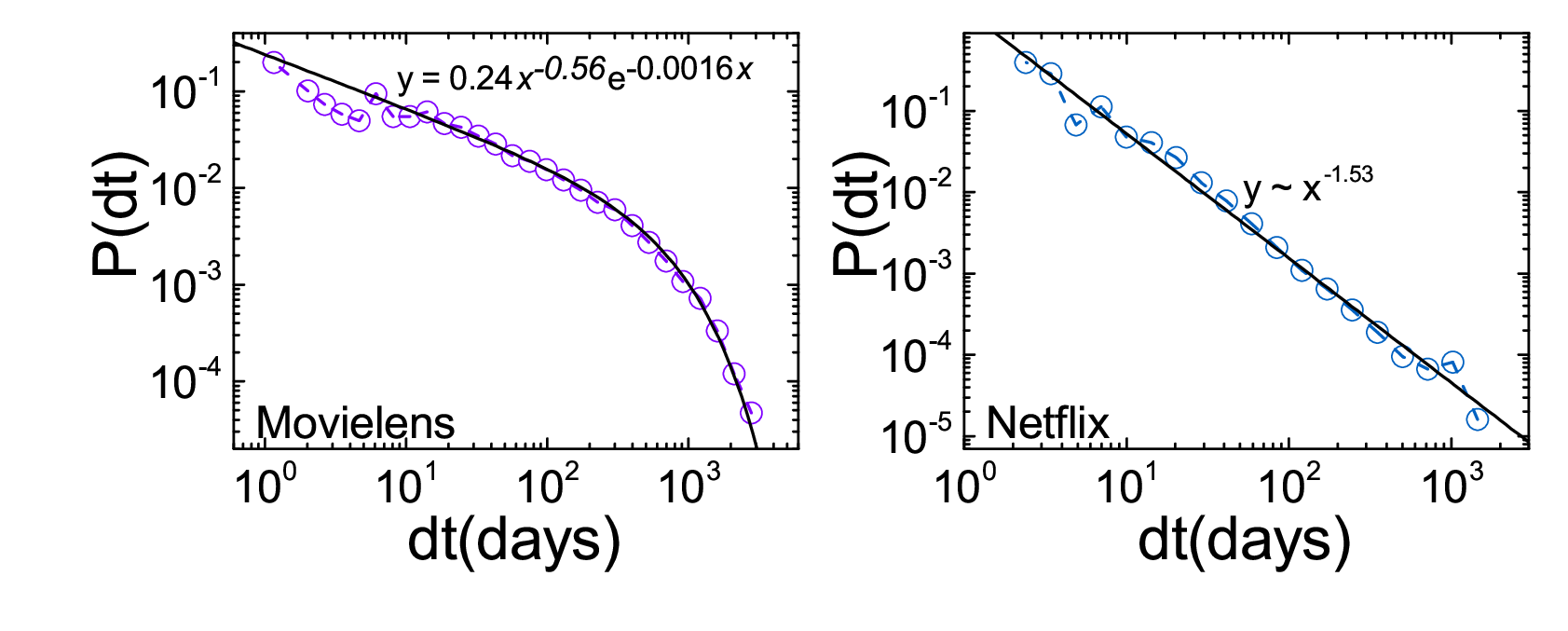}
\caption{\label{fig:timeInter_all}(Color online) Inter-event time
  distribution at the communal level. The power-law with exponential
  cut-off relation behaves in Movielens, while power-law relation
  behaves in Netflix. This result shows the burstiness is generic to the
  system.}
\end{figure}

\begin{figure}[!t]
\centering
\includegraphics[width=0.45\textwidth]{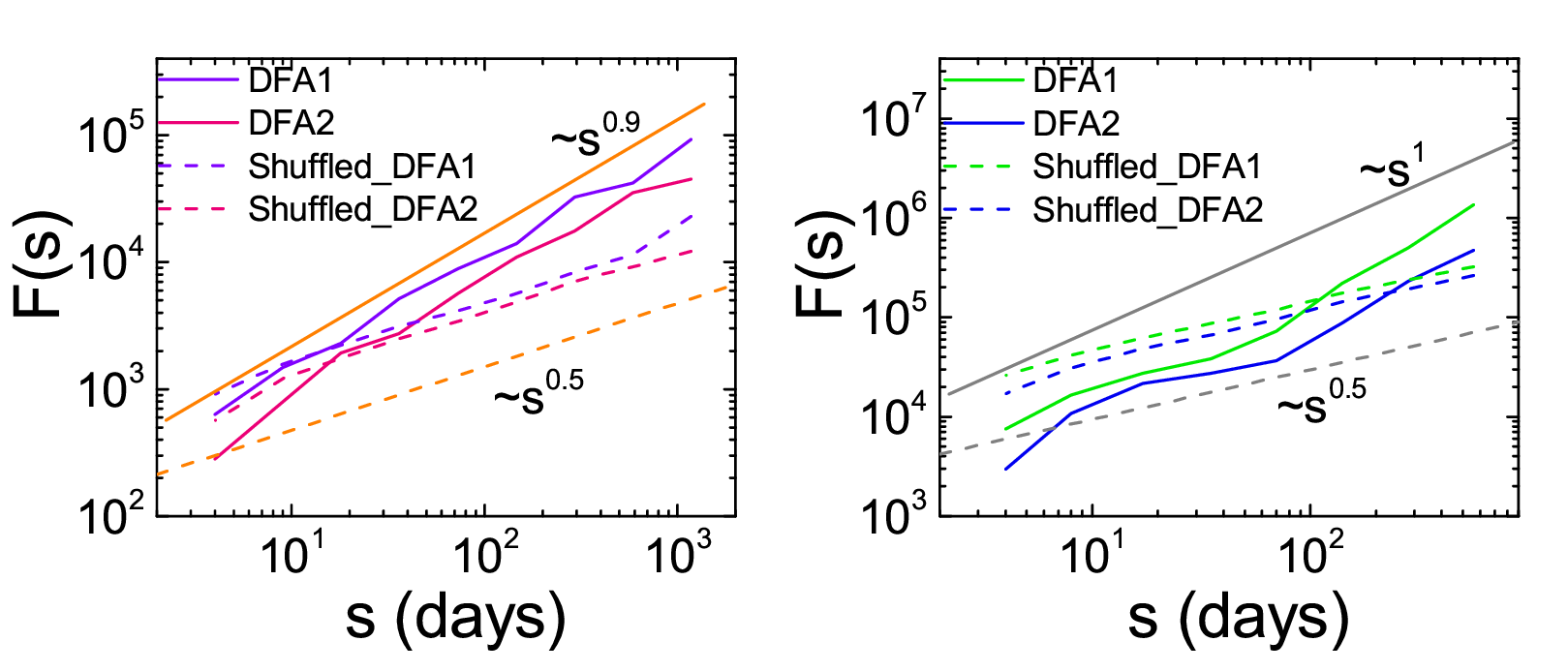}
\caption{\label{fig:DFA_all} (Color online) The results of 1-order and
  2-order DFA for Movielens and Netflix at the commnunal level. The
  Hurst exponents of original time series of records obtained via least
  square estimating method is 0.9 and 1 in Movielens and Netflix
  respectively. When they are randomly shuffled, the Hurst exponents
  approximately reduce to 0.5. This result demonstrates that strongly
  long-term correlations exist in both Movielens and Netflix. Note that solid
  and dash line respectively indicates original time series of records
  and shuffled one.}
\end{figure}

\noindent
The inter-event time distributions with respect to activity levels
have a fat tail, but we still must determine whether this property is
maintained throughout the community (i.e., the entire
system). Figure~\ref{fig:timeInter_all} shows the inter-event time
distributions of Movielens and Netflix at the communal level.
It can be seen that it is fitted by an exponential cut-off
power law distribution for Movielens and approximate power law one for Netflix, which
suggests that the fat tail is generic to the system. Further,
We aggregate the records from all users in the community, and
investigate the resulting time series to quantify the long-term
correlations. Figure~\ref{fig:DFA_all} shows that although the
fluctuation functions are somewhat affected by the oscillations
associated with periodic patterns of activity, the Hurst exponents we
obtain from 1-order and 2-order DFA are robust and approximately 0.9 and
1 for Movielens and Netflix, respectively. They exhibit strong
long-term correlations and are also associated with a time series
spectrum with $\frac{1}{f}$ scaling, suggesting that there is
self-organized criticality in the system. We shuffle these time series
of the entire community to preserve the distribution of inter-event times
and find that the Hurst exponents reduce to 0.5, which
suggests that the long-term systemic correlation is due to interdependence
(which Ref.~\cite{Rybski2012} calls ``true correlation'').


\subsection{Multifractality in community activity}

\noindent
The long-term correlations in online viewing activity of separate individuals and
of the entire community indicate the presence of fractality, but little
research has analyzed the type of fractality involved or its
origin. The results of DFA at the community level fluctuating at double
logarithmic coordination indicates possible multifractality (see
Fig.~\ref{fig:DFA_all}). Inspired by Ref.~\cite{Kantelhardt2002}, we
introduce MFDFA and analyze the datasets to determine whether the
fractality is monofractal or multifractal.

\begin{figure}[!t]
\centering
\includegraphics[width=0.45\textwidth]{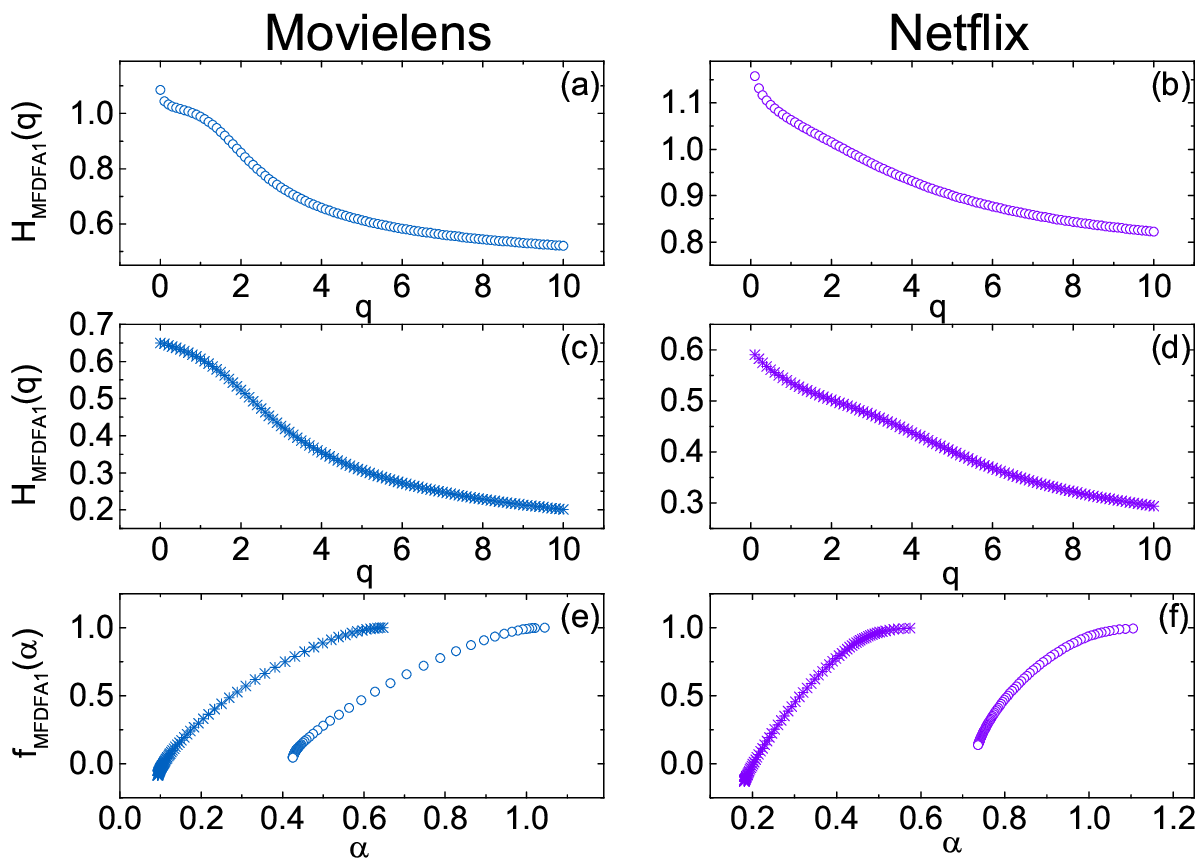}
\caption{\label{fig:MFDFA_real}(Color online) Relation between $H(q)$
  and $q$ deriving from 1-order MFDFA (a)-(d) and the corresponding
  singularity spectrum (e) and (f), where (a) and (b) are obtained from
  the original time series while (c) and (d) are obtained from the
  shuffled ones. Though the multifractality keeps, there are significant
  changes happened to the values of $H$ and $\alpha$. This results
  reveals the existence of multifractality for Netflix and Movielens and
  its formation due to the combined effect of the long-term correlations and the
  broad PDF of inter-event times.}
\end{figure}

Using 1-order MFDFA, we fix a certain value of $q$ and fit $F_q(s)$
and $s$ at double logarithmic coordination with the least square estimating method to
obtain the value of generalized Hurts exponent $H(q)$. Herein, we set $q$ in an interval $(0, 10]$ with
a step length $0.1$. Figure~\ref{fig:MFDFA_real} (a) and (b) show the $H(q)$ as a function of $q$ via
1-order MFDFA for Movielens and Netflix, respectively. We find that both for Movielens and
Netflix $H(q)$ decreases as $q$ increases, i.e., the dependence between $H(q)$ and $q$ suggests
multifractality in community activity.


To determine the origin of such multifractality, we randomly shuffle the
time series of records at the communal level by preserving the
inter-event time distribution and applying 1-order MFDFA
once again on the shuffled one. Figures~\ref{fig:MFDFA_real}(c) and (d) show that
although the $H(q)$ for both Movielens and Netflix are clearly
smaller than the original, the dependence between $H(q)$ and $q$
remains and multifractality is still present.

Much more legible results describing the extent of multifractality in
online viewing activity for Movielens and Netflix are characterized by the
singularity spectrum $f(\alpha)$, as shown in Fig.~\ref{fig:MFDFA_real}(e) and (f).
Note that the horizon span of $f(\alpha)$ both for the original and the shuffled
time series are trivially different, which is suggested by the difference of the asymptotical
values of $H(q)$, $\Delta \alpha=1.38$ (original) and $\Delta \alpha=1.05$ (shuffled)
for Movielens and $\Delta \alpha=0.78$ (original) and $\Delta \alpha=0.64$ (shuffled) for
Netflix. And the more large changes happen to the values of $\alpha$.
We derive these results using the relation between $H(q)$ and $q$ described in \cite{Kantelhardt2003,Kantelhardt2006},
It confirms the dependence between $H(q)$ and $q$, and also suggests that
the multifractality in community activity is not solely induced by the
long-term correlations (see in Appendix C)



\begin{figure}[!t]
\centering
\includegraphics[width=0.45\textwidth]{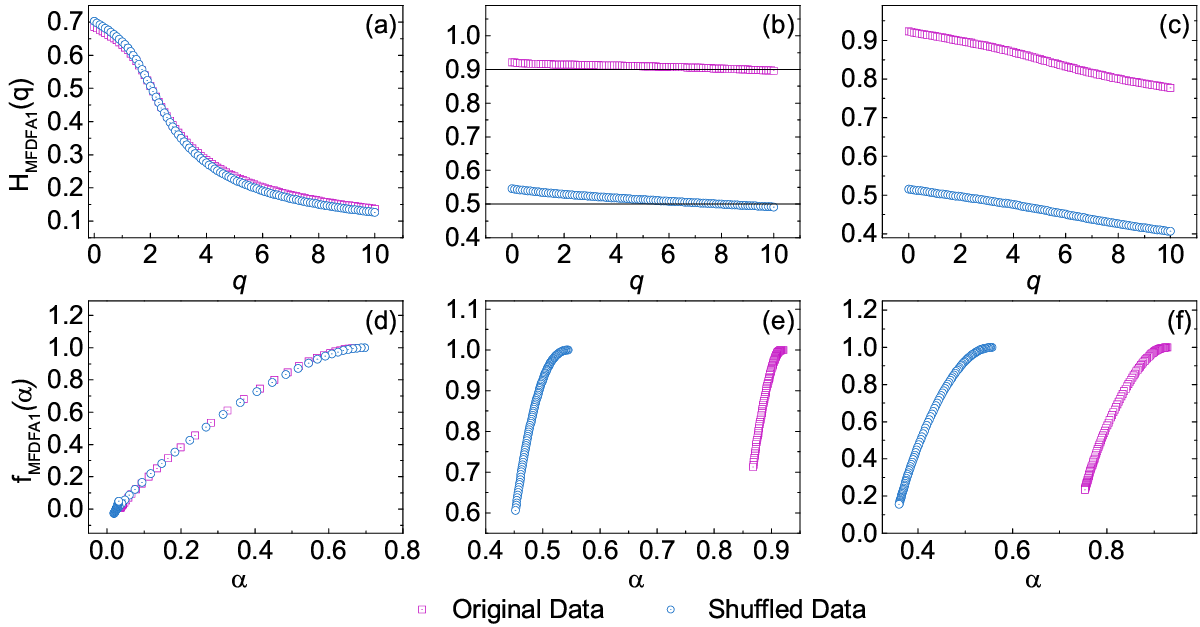}
\caption{\label{fig:MFDFA_synthesis} (Color online) Relation between
  $H(q)$ and $q$ obtained from 1-order MFDFA (a)-(c) and the
  corresponding singularity spectrum (d)-(f) of three types of synthetic
  time series, where pink square and blue circle respectively indicates
  the results of original and shuffled time series. Note that the first
  column shows a synthetic time series obeying a power law distribution
  $y \sim x^{-2}$, the second one describes a synthetic time series
  whose hurst exponent is 0.9, the third one represents a synthetic time
  series that combines the properties of the former two time series.
  Through carefully analyzing them, we can find that only the third time
  series behaves similar results to empirical findings.}
\end{figure}

Our analysis leads us to assume that the multifractality in online
viewing activity from Netflix and Movielens is induced by the combined effect of
the long-term correlations and the broad PDF of inter-event times. To verify our hypothesis, we
analyze the multifractality of three synthetic time series that are
analogous to real time series. The first is a random series that obeys a
power-law distribution ($y \sim x^{-2}$), the second is a monofractal
series with strong long-term correlations ($H=0.9$), and the third is a
combination of the first two (see Appendix A). We also obtain the
corresponding shuffled time series and use MFDFA to derive the
multifractality (see Fig.~\ref{fig:MFDFA_synthesis}).

Figures~\ref{fig:MFDFA_synthesis}(a) and (d) show that in the
random series that obeys a power-law distribution there
is a remarkable dependence between $H(q)$ and $q$ (there is a broad
singularity spectrum), which indicates multifractality dominated by the
power-law distribution. They also show that the absence of long-term
correlations causes an overlap in results between the original and
shuffled time series. Figures~\ref{fig:MFDFA_synthesis}(b) and (e)
show that $H(q)$ and $q$ are independent (there is a narrow singularity spectrum),
which indicates the monofractility in the monofractal series with strong long-term
correlations. The long-term correlations cause $H(q)$ to change from 0.9
to 0.5. Figures~\ref{fig:MFDFA_synthesis}(c) and (f) show that in the time series that combines
the other two the significant horizon span of singularity spectrum and
change of $H(q)$ produces results that are similar to empirical
findings. Our analysis strongly indicates that the multifractality in
online viewing activity is caused by the broad PDF of inter-event times and the presence
of long-term correlations.


\section{Conclusion}

\noindent
We have analyzed the datasets of online viewing activity from Movielens and Netflix
at both the individual and communal level. At the individual level we
find fat-tailed inter-event time distributions and the dependence on
long-term correlations. Our analytical results are not exact, as in
Ref.~\cite{Rybski2012}, because the inter-event
time distributions are restricted to the activity levels
and not strictly power law. At the communal level
we find properties that are similar to
those at the individual level, but here the long-term correlations are
caused by the interdependence of community activity. Furthermore, the long-term
correlations characterized by the Hurst exponent derived from DFA imply
the presence of fractality in online viewing activity.

To determine the type of such fractality and its origin, we apply
MFDFA and find multifractality at the communal level. We hypothesize
that this is caused by the combined effect of the broad PDF of inter-event times and
the long-term correlations. We verify this by analyzing three types of
synthetic time series that have at least property in common with a real
time series. Thus, we can conclude that a dual-induced multifractality
exists in online viewing activity, which enlarges this generic property
commonly found in human activity from physical space to cyberspace.
Nevertheless, it shouldn¡¯t be ignored that an appropriate model lacking to explain
the mechanism of reproducing such time series. According to \cite{Ashkenazy2003,Schmitt2009,Extremera2016},
the time series of online viewing activity can be decomposed into magnitude and sigh time series,
and by systematically analyzing them we may obtain more dynamical properties to explain
the mechanism of multifractality in online viewing activity. We hope that future work will solve these problems.

\section*{Acknowledge}

\noindent
We thank the financial support from the National Natural Science
Foundation of China (Grant Nos. 61673086, 61603074, 71571017, 91646124 and 71621001).
The Center for Polymer Studies of Boston University is supported by NSF
Grants PHY-1505000, CMMI-1125290, and CHE-1213217, by DTRA Grant
HDTRA1-14-1-0017, and by DOE Contract DE-AC07-05Id14517.

\section*{Appendix}
\setcounter{figure}{0}
\renewcommand\thefigure{A\arabic{figure}}

\subsection{Construction of synthetic time series}

\noindent
To construct the three synthetic time series, we first synthesize a
random time series $x(t)$ that obeys a power-law distribution $p(x)=
\beta x^{-(1+\beta)}$.  We use the central limit theorem and generate it
to be
\begin{equation}
x(t) = (r(t)/\beta)^{-\frac{1}{1+\beta}},
\end{equation}
where $r(t)$ is a time series sampled from a uniform distribution
$U(0,1)$. Here we set $\beta=1$, which causes $x(t)$ to obey a power-law
distribution $p(x) = x^{-2}$.

We then apply the Flourier filtering method proposed in
Ref.~\cite{Makse1996} to generate a monofractal time series with
long-term correlations. The procedure is as follows:

\begin{itemize}

\item[{(i)}] Generate a 1-dimensional random time series $U_i$ that
  follows a Gaussian distribution, and derive its Fourier transform
  coefficients $U_q$.

\item[{(ii)}] Obtain $S_q$ from the Fourier transformation of $C_l$,
  where $C_l= \langle \mu_i \mu_{i+l}\rangle= (1+l^2)^{-\gamma/2}$.

\item[{(iii)}] Calculate $N_q=[S_q]^{1/2} U_q$.

\item[{(iv)}] Derive the time series $N_r$ using the inverse Fourier
  transformation of $N_q$. In this way we transform $N_r$ using
  $N_r=N_r- min(N_r)+1$.

\end{itemize}

We combine these two time series and synthesize the third time series,
\begin{equation}
X(t)=(N_r(t)/\beta)^{-\frac{1}{1+\beta}},
\end{equation}
where $N_r$ is the time series with long-term correlations.


\subsection{$F$-test for linear regression}

\begin{figure}[!t]
\centering
\includegraphics[width=0.45\textwidth]{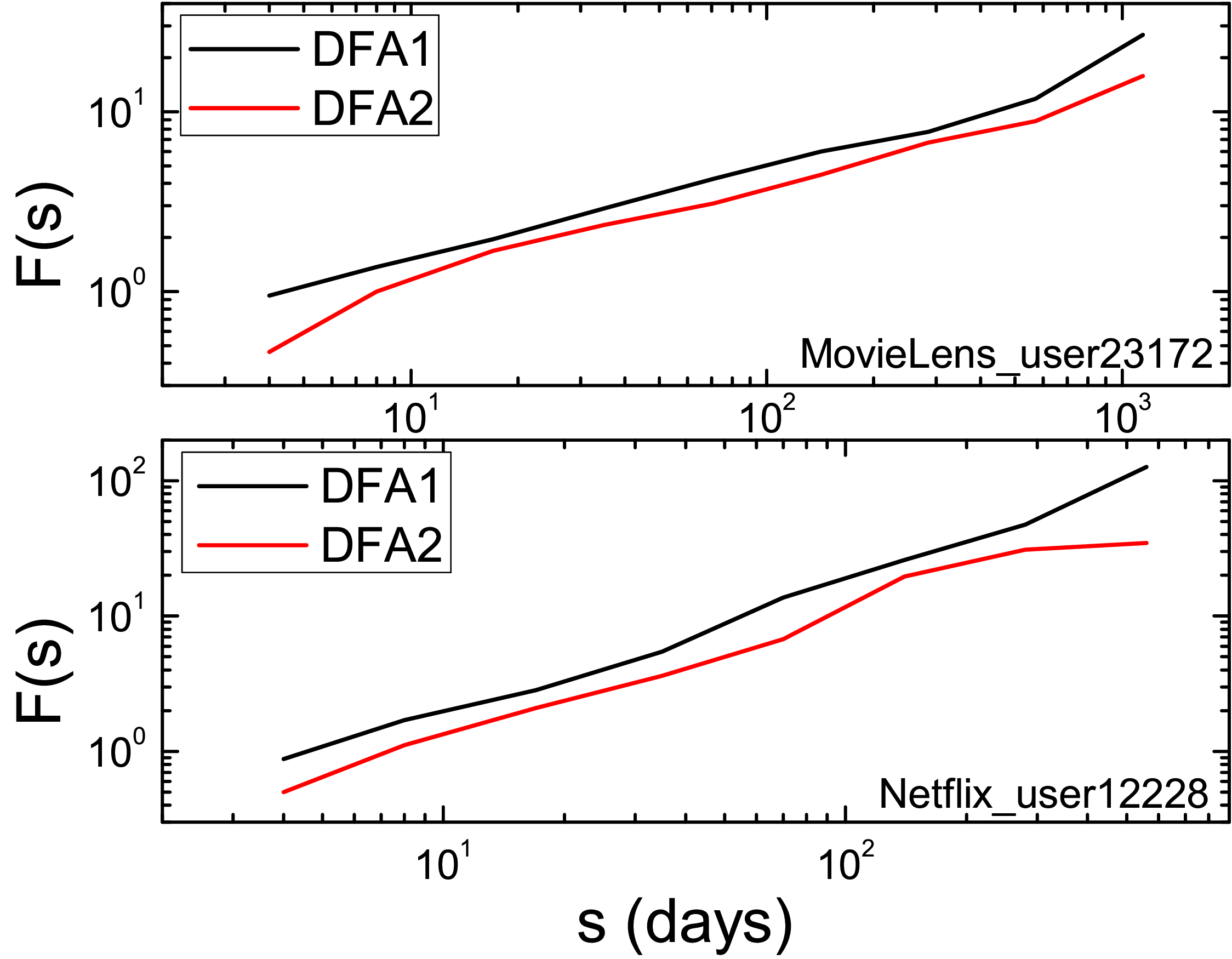}
\caption{\label{fig:A0}(Color online) The fluctuation functions of DFA at a log-log scale from two uesrs.
They clearly show the power-law relation across multiple scales for both Movielens and Netflix.
And for each user, the power-law relations obtained from DFA-1 and DFA-2 show a approximately
same trend across multiple scales, which suggest that the scaling exponents of long-term correlations
are not sensitive to the order of DFA.}
\end{figure}

\noindent
We firstly show the fluctuation functions of DFA at a log-log scale from two uesrs.
Figure \ref{fig:A0} shows the power-law relations of fluctuation functions across multiple scales
for both Movielens and Netflix. We estimate the scale exponents of long-term correlations via
fitting linear relation between $\log(F(s))$ and $\log(s)$. Then,
we test the linear relation to determine whether the linear relation
between two variables $x$ and $y$ is significant (i.e.,
$y=\beta_{0}+\beta_{1}x+\epsilon$). We do this by testing the following
hypothesis and its alternative:

$H_0$:$\beta_1=0$, the relation between these two variables is not significant.

$H_1$:$\beta_1\neq0$, the relation between these two variables is significant.

In the $F$-test we introduce $F$ to measure the strength of the
connection between $x$ and $y$, which is defined
\begin{equation}
F = \frac{\sum (\widehat{y_i} - \overline{y})^2}{\frac{1}{n - 2} \sum
  (y_{i} - \widehat{y_{i}})^2} \sim F(1,n - 2),
\end{equation}
where $\widehat{y_{i}}$ is derived from the equation
$y=\beta_{0}+\beta_{1}x$, $\overline{y}$ is the mean value of $y_i$, and
$x_i$ and $y_i$ are the real data.


\begin{figure}[!t]
\centering
\includegraphics[width=0.45\textwidth]{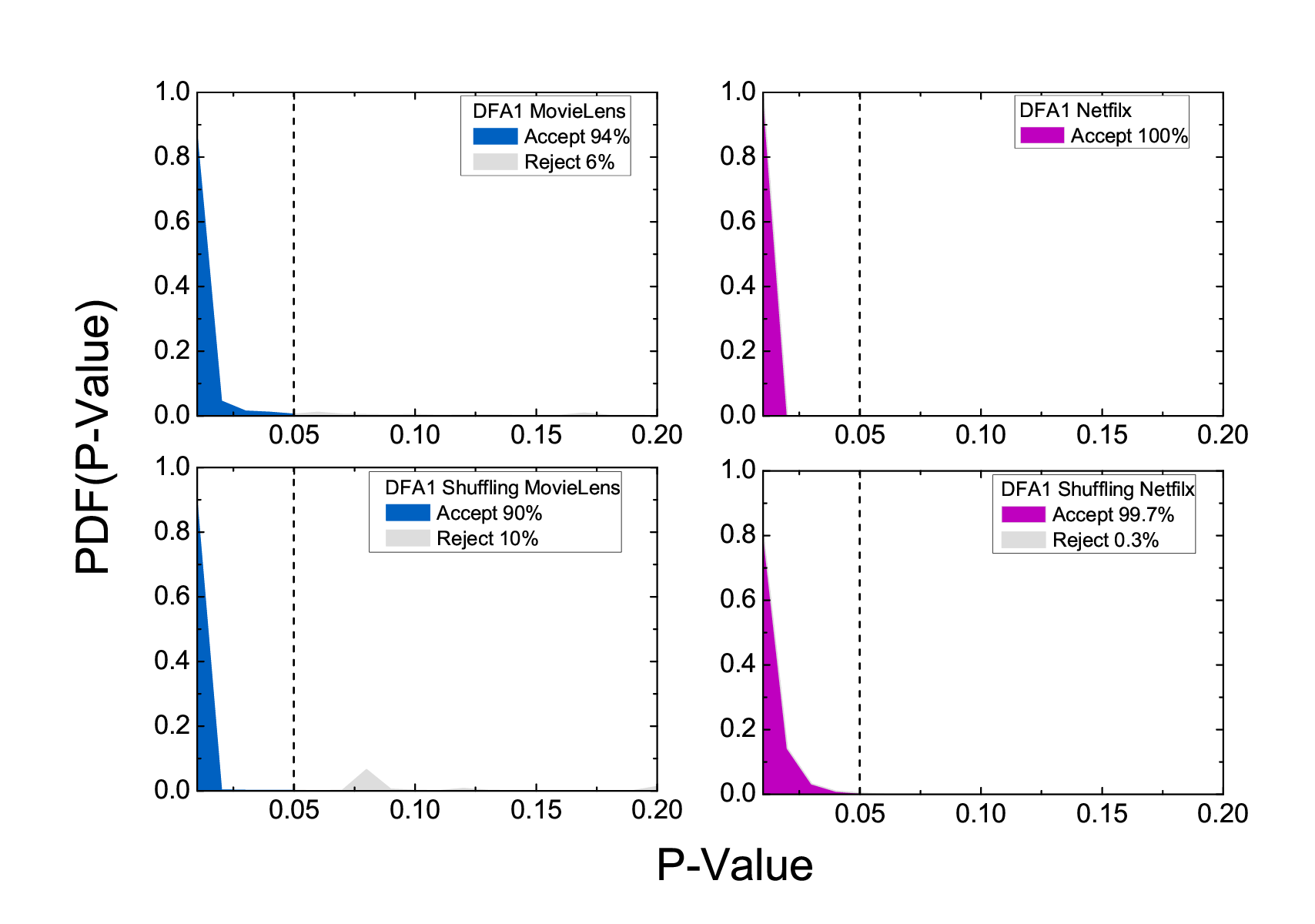}
\caption{\label{fig:A1}(Color online) The distribution of users'
  $P$-values for original data and corresponding shuffled one. It shows
  that the strongly linear correlation between $\log(F(s))$ and
  $\log(s)$ for most users, which demonstrates the reliability of the
  results in Fig.4.}
\end{figure}

Using each $F$-value, we can derive the $P$-value for the distribution
$F(1,n - 2)$. At confidence level $\alpha$ (here $\alpha=0.05$), when
$P<\alpha$ we accept $H_1$ and thus accept the linear relation. When
$P>\alpha$ we reject $H_1$ and thus reject the linear relation. We
compute the $P$ value for each user to fit the relation between
$\log(F(s))$ and $\log(s)$ derived from the DFA. Figure~A2 shows the $P$-value
distribution that indicates a strong linear correlation between
$\log(F(s))$ and $\log(s)$ for most users in both the original data and
the shuffled data. This confirms the reliability of the results in shown
in Fig.~\ref{fig:DFA_level}.

\subsection{Surrogate methods for analyzing multifractality}


\begin{figure}[!t]
\centering
\includegraphics[width=0.45\textwidth]{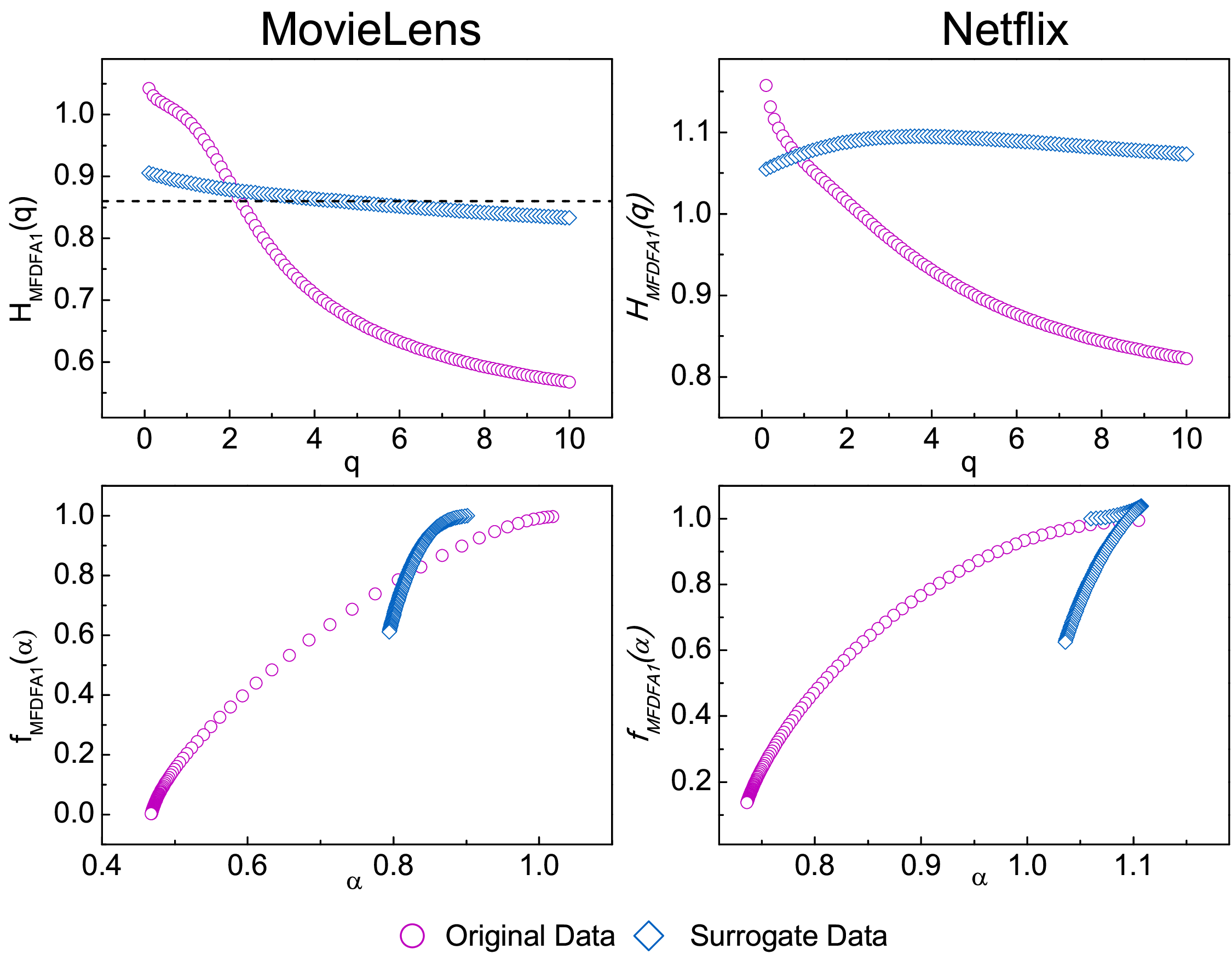}
\caption{\label{fig:A3}(Color online) Relation between $H(q)$ and $q$
  deriving from 1-order MFDFA and the corresponding singularity
  spectrum. The pink circle indicates the original data, while the blue
  square represents the surrogate data produced by \emph{only} Fourier
  phase randomization. When the broad PDF of inter-event times is disrupted, the long-term
  correlations still lies in the surrogate data but the multifractality
  is dramatically weaken.}
\end{figure}


\begin{figure}[!t]
\centering
\includegraphics[width=0.45\textwidth]{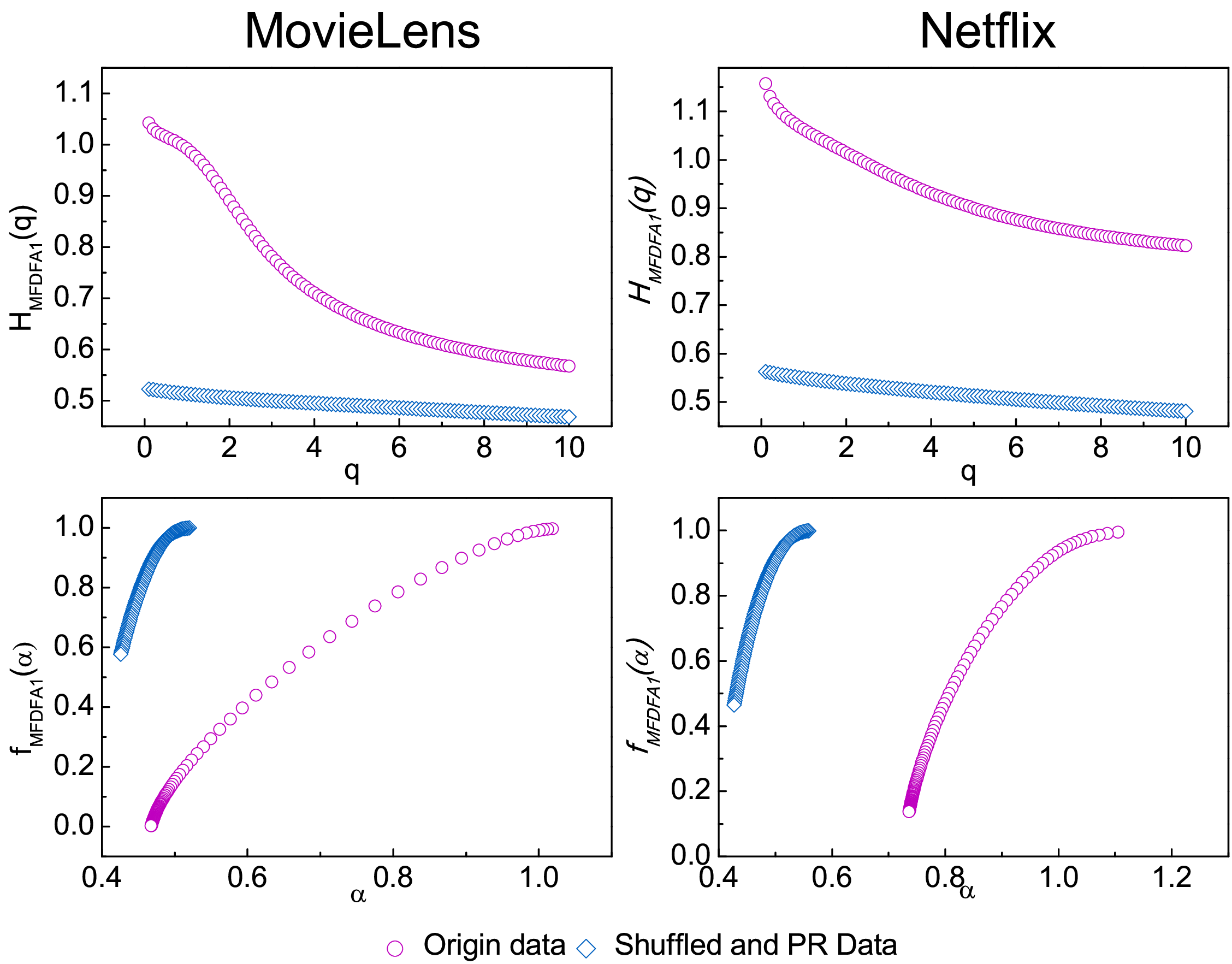}
\caption{\label{fig:A4}(Color online) Relation between $H(q)$ and $q$
  deriving from 1-order MFDFA and the corresponding singularity
  spectrum. The pink circle indicates the original data, while the blue
  square represents the surrogate data produced by \emph{both} shuffling
  and Fourier phase randomization. When the broad PDF of inter-event times and long-term
  correlations are both missed, $H(q)$ is close to $0.5$ and has a
  trivial relation with $q$. Moreover, the singularity spectrum also has
  a very narrow horizon span.}
\end{figure}

As mentioned above, there are two key factors that affect the
multifractality in a time series of records: (i) long-term correlations
in the fluctuations and (ii) a broad PDF of inter-event times. When we analyze the
contributions of these two factors affecting multifratility separately,
we generate many surrogate time series through shuffling and phase
randomization \cite{Norouzzadeh}. The shuffling procedure preserves the
PDF of the time series of records but destroys any long-term
correlations. Thus we randomly sort the entire time series at least 10
times. Fourier phase randomization maintains the long-term correlations
but disrupts the broad PDF of inter-event times \cite{Theiler,Prichard}.

Figures A3 and A4 show multifractality generated by shuffling and
Fourier phase randomization. Figure~A3 shows that when using only
Fourier phase randomization to destroy the broad PDF of inter-event times, the long-term
correlations remain in the surrogate data but the multifractality is
weakened. Figure~A4 shows that when applying both shuffling and Fourier
phase randomization to reduce long-term correlations and destroy the
broad PDF of inter-event times, the $H(q)$ value is close to $0.5$ and has a trivial relation
with $q$, and the horizon of the singularity spectrum is narrowed.
These results verify the presence of dual-induced multifractality in
online viewing activity.

\section{Reference}

\providecommand{\newblock}{}

\end{document}